\documentclass[12pt]{iopart}
\usepackage{graphicx}
\usepackage{iopams}
\expandafter\let\csname equation*\endcsname\relax
\expandafter\let\csname endequation*\endcsname\relax
\usepackage{amsmath,amsfonts,amsthm,bm,mathtools}
\usepackage[usenames,dvipsnames]{color}
\usepackage{url}
\usepackage{epstopdf}

\usepackage{cite}

\usepackage{pdfpages}

\definecolor{linkcolor}{rgb}{0,0,0.6} 
\usepackage[colorlinks=true, pdfstartview=FitV, linkcolor= linkcolor, citecolor= linkcolor, urlcolor= linkcolor, hyperindex=true,hyperfigures=true]{hyperref}

\renewcommand{\O}{{R}}
\newcommand{\trap}{{\mathrm{trap}}}
\newcommand{\diff}{{\mathrm{diff}}}
\newcommand{\tot}{{\mathrm{tot}}}
\newcommand{\sys}{{\mathrm{sys}}}
\newcommand{\med}{{\mathrm{med}}}
\newcommand{\KL}{{\mathbb S}}
\newcommand{\SC}{{\kappa}}
\newcommand{\fric}{\widehat{\gamma}}
\newcommand{\Eul}{{G}}

\begin{document}

\title[]{A thermodynamic uncertainty relation for a system with memory}

\author{Ivan Di Terlizzi$^{*,\dag}$}
\author{Marco Baiesi$^{*,\dag}$}

\address{$*$ Dipartimento di Fisica e Astronomia ``Galileo Galilei'',
 Universit\`a di Padova, Via Marzolo 8, 35131, Padova, Italy
} 
\address{$\dag$ INFN, Sezione di Padova, Via Marzolo 8, 35131, Padova,
Italy}


\begin{abstract}
We introduce an example of thermodynamic uncertainty relation (TUR) for systems modeled by a one-dimensional generalised Langevin dynamics with memory, determining the motion of a micro-bead driven in a complex fluid. Contrary to TURs typically discussed in the previous years, our observables and the entropy production rate are one-time variables. The bound to the signal-to-noise ratio of such state-dependent observables only in some cases can be mapped to the entropy production rate. For example, this is true in Markovian systems. Hence, the presence of memory in the system complicates the thermodynamic interpretation of the uncertainty relation.
\end{abstract}


\noindent{\it Keywords\/}: nonequilibrium inequalities, memory effects, diffusion, fluctuations, entropy production

\maketitle

\section{Introduction}
The performances of non-deterministic systems are intrinsically limited by several laws.
For instance, the {\em quantum speed limit} determines the minimum time needed to transform a quantum state to another~\cite{def17}. Recently is was noted that similar speed limits exist also in the evolution of stochastic classical systems~\cite{oku18,sha18,shi18,fun19}.
Other classical bounds limit the statistics of currents and other observables, whose squared average must be lower that their variance times some cost function~\cite{hor20}. The {\em thermodynamic uncertainty relation} (TUR)~\cite{bar15,sei18fpsp} is the primary example of such nonequilibrium inequalities and includes the entropy production as a cost function~\cite{pie16,pol16,gin16,nar17,pie17,gin17,hor17,pro17,mac18,dec18,ito18arx,tim19,bus19,dec19,dec20,koy20} (dissipation may also limit the speed of a process~\cite{fal20arx}). The TUR and its generalisations~\cite{pie16a,mae17,gar17,chi18,dit19,vu19inert,bar19} are inequalities usually discussed and proven for discrete and continuous diffusive Markov systems. Fewer results are available for non-Markovian systems~\cite{vu19,vu20,pot19}, namely for systems with some form of memory.

On the practical side, TURs may help in thermodynamic inference~\cite{sei19}, for instance in evaluating the entropy production rate from data~\cite{gin17,li19,man20,ots20arx,vu20est}.
Theoretically, while the proof of the TUR in steady states is provided by the machinery of large deviation theory~\cite{gin16,hor17}, an approach by Dechant and Sasa (DS)~\cite{dec18,dec20} adopts information theory as the main theoretical tool. Moreover, a unifying view~\cite{fal19} may explain the mechanism behind uncertainty relations.

The DS approach leads to quite general results for stochastic systems, finite times statistics, and regimes outside steady states. For instance, DS showed that various forms of TUR hold for diffusion processes and anticipated that a TUR holds also for Langevin dynamics with inertia. Following this approach, Van Vu and Hasegawa~\cite{vu19inert} indeed provided an explicit nonequilibrium inequality for inertial stochastic dynamics. Their formula shows that the mean entropy production cannot by itself constitute the cost function in inertial systems. In equilibrium, where on average dissipation is absent while currents are eventually present thanks to inertia, there is a form of dynamical activity that naturally enters in the upper bound, as in a {\em kinetic uncertainty relation}~\cite{dit19} or similar inequalities including time-symmetric nondissipative observables~\cite{pie16a,mae17,gar17,chi18,vu19inert}.

A linear response formula was used by DS to compares the probability of observing an event for two different processes~\cite{dec20}. Usually, in nonequilibrium statistical physics, the comparison has been done between path weights of two kinds of dynamics (see for instance the derivation of linear response formulas in~\cite{bai09,bai09b,bai10}). However, one may also consider the instantaneous probability of observables as position and velocity~\cite{ito18arx,dec20}. We are going to follow this path.

In this work we provide a nonequilibrium inequality for a simple system with memory. This is achieved by developing the DS approach for a one-dimensional Langevin system subject to a friction dynamics with memory and to a time dependent force. A practical realization of this dynamics is a colloidal bead dragged by optical tweezers in a complex medium~\cite{gom11prl,berner2018oscillating}, say a viscoelastic fluid, or driven by a space-independent time-modulated force. Analytical solutions show that the average and variance of observables for this non-Markovian system also obey a form of generalized TUR. In this TUR there appear instantaneous quantities: the instantaneous rate of entropy production is present, while original TURs include the accumulated entropy production and may reduce to forms including its rate only if a steady state is established.  Similarly, the observable $\O$ entering in our formula is state dependent and its variance is meant as the instantaneous variance in a statistical ensemble at the same time.
Thus, the formulas in this work are not for integrated currents, as in usual TURs, but for state-dependent quantities. The importance of such instantaneous quantities was recently highlighted in a novel version of TUR for Markovian systems~\cite{koy20} (see also previous examples~\cite{ito18arx}).

For the simplest Langevin dynamics without memory (that is, with friction depending on the instantaneous velocity and with white noise) and with harmonic trap, both in steady states (where the trap has been moving with a steady velocity for a long time) and in transients from an initial equilibrium, we find that the entropy production rate is the only component of the cost function in the TUR. However, it comes with a prefactor depending on the trap strength and the fluid viscosity.
More in general, for long times, a thermodynamic interpretation recovers the entropy production rate as the asymptotically relevant part of the cost function.

Our calculations are based on the modified Laplace transforms that we recently introduced~\cite{OurGLE}. In order to deal with steady states, in this modification the initial time is shifted earlier than time zero, at which we start the statistical analysis. By sending the initial time to minus infinity, we may well describe systems where the memory effects are perfectly developed.

In the following section we introduce the model and some of its features. The main formulas are discussed in Section~\ref{sec:main} and their applications can be found in Section~\ref{sec:appl}. After our conclusions in Section~\ref{sec:concl}, some technical details are discussed in a set of Appendices.

\section{Model}
\label{sec:model}

We study a Langevin dynamics with colored noise and friction with memory. We would like to characterize uncertainty relations for the motion of a colloidal particle in a complex viscoelastic fluid, subject to a harmonic potential due to the action of optical tweezers, and eventually to a time-modulated space-independent external force $f(t)$ that could represent a uniform electric field acting on a charged particle.
A simpler Langevin equation with white noise and Markovian dynamics would not be suitable for modeling memory effects by the complex fluid. By introducing a memory kernel $\Gamma(t)$, we thus consider the generalized Langevin equation (GLE)
\begin{equation} \label{eq1}
 m\ddot x(t) = - \int_{t_{m}}^{t}\Gamma(t-t') \dot x(t') \mathrm{d}t'- \SC \left[ x (t) - \lambda(t)\right] + f(t) + \eta (t)
\end{equation} 
where $\SC$ is the spring constant associated to a harmonic trap with time dependent minimum $\lambda(t)$ and $t_{m}\le 0$ is the time to which the effects of memory extend. Moreover, $m$ is the mass of the colloidal bead and $\eta (t)$ is a coloured Gaussian noise such that $\langle \eta (t) \rangle = 0$ and $\langle \eta (t) \eta \left( s \right) \rangle = k_{B} T \Gamma \left( |t -s| \right)$ obeying the fluctuation-dissipation theorem~\cite{kub66}. For future convenience, we collect the space-independent terms in
\begin{equation}
F(t) \equiv f(t)+\SC \lambda (t)
\end{equation}
A trajectory thus starts from an initial condition $(x_{t_{m}},v_{t_{m}})$ and evolves according to \eqref{eq1}. The initial time $t_m\le 0$ is finite. However, the limit $t_m \to -\infty$ will be useful for describing steady regimes when, for example, the harmonic trap is moving at constant velocity and no other external force is present.

In this paper we will only focus on processes for which the position probability density function (PDF) at time $t$ is Gaussian. For linear systems, as in our case, this can be obtained either by starting already from a Gaussian PDF $P(x_{t_m},v_{t_m},t_m)$, or by starting from an arbitrary distribution and wait long enough till it becomes a Gaussian. The latter scenario occurs if either $t_m \to -\infty$ or $t \to \infty$. Under these hypothesis, the PDF $P(x_{t},v_{t},t|*)$, were $*$ stands for the initial conditions discussed above, is a bivariate Gaussian and by marginalising with respect to the velocity $v_{t}$ we obtain a Gaussian PDF for the position,
\begin{equation}
  P \left( x , t ,*\right) = \frac{1}{\sqrt{2 \pi \langle \Delta^{2} x \rangle_{*,t}} } \exp \bigg[ - \frac{\left( x - \langle x \rangle_{*,t} \right)^{2}}{2 \langle \Delta^{2} x \rangle_{*,t} } \bigg]
\end{equation}
This distribution is of course completely characterised by the average position $\langle x \rangle_{*,t}$ and its variance $\langle \Delta^{2} x \rangle_{*,t}$ at time $t$. Both can be calculated starting from \eqref{eq1} by applying a {\it modified} Laplace transform, defined as
\begin{equation}
  \hat{g}_{t_{m}}(k) = \mathcal{L}^{t_{m}}_{k} [g(t)] \equiv \int_{t_{m}}^{+\infty} \e^{-kt}g(t)\mathrm{d}t
\end{equation}
Its details are discussed in~\cite{OurGLE}.
The sub/superscript $t_m$ is useful for reminding us that this transform is different from the standard Laplace transform. For a causal function, $g(t)=0$ if $t<0$, there is no difference between the usual Laplace transform and the modified one. However, this is not the case in general.

An important quantity that appears while solving the GLE is the "position susceptibility" $\chi_{x}(t)$, defined via its modified Laplace transform
\begin{equation}
\hat{\chi}_{x}(k) = [m k^{2}+k\hat{\Gamma}(k)+\SC]^{-1}
\end{equation}
In the following we will also use its integral $\chi(t)$ and its derivative $\chi_{v}(t)$ (velocity susceptibility),
\begin{align} \label{chi}
  \chi(t)    &\equiv \int_{0}^{t}\chi_{x}(t') \mathrm{d}t'\\
  \chi_{v}(t) &\equiv \partial_{t} \chi_{x}(t)
\end{align}
With these definitions and following the procedure in~\cite{OurGLE}, it can be shown  that the average position is equal to
\begin{equation} \label{eq6}
\begin{split}
  & \langle x \rangle_{v_{t_m},x_{t_{m}},t} = \langle x_{t_m} \rangle(1-\SC \chi (t-t_{m}))+ m\langle v_{t_{m}}\rangle \chi_{x}(t-t_m) + \int^{t}_{t_{m}} \chi_{x} (t-t') F(t') \mathrm{d}t' \\
\end{split}
\end{equation}
where $\langle x_{t_m} \rangle$ and $\langle v_{t_m} \rangle$ are respectively the average position and velocity at the initial time $t_m$. This also explains terming $\chi_{x}(t)$ a susceptibility. We are using the fact that the presence of the the external force $f(t)$ does not change the results obtained in \cite{OurGLE}. For the variance of the position one obtains
\begin{equation} \label{eq7}
\begin{split}
 \langle \Delta^{2} x \rangle_{v_{t_m},x_{t_m},t} =& k_{B} T \Big[ 2 \chi(t-t_m) - m \chi^{2}_{x}(t-t_m) - \SC \chi^{2}(t-t_{m}) \Big] +\\
  &+ \langle \Delta^{2}x_{t_m}\rangle (1-\SC \chi (t-t_{m}))^{2}+ m^{2}\langle \Delta^{2} v_{t_{m}}\rangle \chi^{2}_{x}(t-t_m) + \\
  & + 2m \textrm{Cov}(x_{t_m},v_{t_m})\chi_{x}(t-t_m)(1-\SC \chi (t-t_{m}))
\end{split}
\end{equation}
with initial variances $\langle \Delta^{2}x_{t_m}\rangle$, $\langle \Delta^{2}v_{t_m}\rangle$ and covariance $\textrm{Cov}(x_{t_m},v_{t_m})$.

\section{General result}
\label{sec:main}

In this paper we will use the DS approach~\cite{dec20} (see below) to obtain new stochastic inequalities involving average and variance of a generic position-dependent observable $\O (x)$. This is done by performing a perturbation dependent on a small parameter $\alpha$ such that $P \left( x , t |*\right) \to P^{\alpha} \left( x , t |*\right)$ and $\langle \O \rangle_{*,t} = \int \mathrm{d}x \O \left(x \right) P \left( x , t |*\right) \to \langle \O \rangle^{\alpha}_{*,t} = \int \mathrm{d}x \O \left(x \right) P^{\alpha} \left( x , t |*\right)$. For $\alpha\approx 0$, to leading order it holds
\begin{equation} \label{eq9} 
   \frac{\Big( \big \langle \O \big \rangle_{*,t}^{\alpha} - \big \langle \O \big \rangle_{*,t} \Big)^{2}}{ \big \langle \Delta^{2} \O \big \rangle_{*,t}} \le 2\KL_{t}( t |*)
\end{equation}
where there appears the Kullback-Leibler (KL) divergence between $P$ and $P^{\alpha}$,
\begin{equation}
  \KL_{t}( t |*) \equiv \int \mathrm{d} x P^{\alpha} \left( x, t|* \right) \ln \left[ \frac{ P^{\alpha} \left( x, t | *\right)}{ P \left( x, t|* \right)}\right]
\end{equation}
As in a previous study~\cite{dit19}, we now choose an $\alpha$-dependent perturbation that maps to an ensemble at a time rescaled by $1+ \alpha$, namely
\begin{equation}
  P^{\alpha} ( x,t |*) = P \left( x,\left( 1 + \alpha \right) t |* \right) \implies \langle \O \rangle^{\alpha}_{*,t} = \langle \O \rangle_{*,\left( 1 + \alpha \right)t} \approx \langle \O \rangle_{*,t} + \alpha t \langle \dot{\O} \rangle_{*,t}
\end{equation}
so that equation \eqref{eq9} becomes (neglecting orders higher than $\alpha^{2}$)
\begin{equation} \label{eq12}
    \frac{\left(\alpha t\big \langle \dot{\O} \big \rangle_{*,t}\right)^{2}}{\big \langle \Delta^{2} \O \big \rangle_{*,t}} \le 2\KL_{t}( t |*)
\end{equation}
Further calculations, detailed in~\ref{sec:AppB}, also show that the KL divergence for a Gaussian PDF becomes 
\begin{equation} \label{eq13}
   \KL( t |*) = \frac{\alpha^{2} t^{2}}{2} \left[ \frac{1}{2} \left(\frac{\partial_t\langle \Delta^{2} x \rangle_{*,t}}{\langle \Delta^{2} x \rangle_{*,t}}\right)^2 + \frac{\left( \partial_t\langle x \rangle_{*,t} \right)^{2}}{\langle \Delta^{2} x \rangle_{*,t}} \right] + \mathcal{O} \left( \alpha^{3}\right) \approx \frac{\alpha^{2} t^{2}}{2} \mathcal{I}(t |*) 
\end{equation}
where $\mathcal{I}(t |*)$ is the Fisher information, i.e., the concavity of the Kullback-Leibler divergence evaluated at its minimum. This allows us to rewrite \eqref{eq12} as 
\begin{equation} 
    \frac{ \big \langle \dot{\O} \big \rangle_{*,t}^{2}}{\big \langle \Delta^{2} \O \big \rangle_{*,t}} \le \mathcal{I}( t |*)
\end{equation}
that is a form of the generalised Cramer-Rao bound, or more explicitly  
\begin{equation} \label{eq:main1}
\begin{split}
&\hspace{2cm} \frac{\big \langle \dot{\O} \big \rangle_{*,t}^{2}}{\big \langle \Delta^{2} \O \big \rangle_{*,t}} \le \frac{1}{2} \left(\frac{\partial_t\langle \Delta^{2} x \rangle_{*,t}}{\langle \Delta^{2} x \rangle_{*,t}}\right)^2 + \frac{\left( \partial_t\langle x \rangle_{*,t} \right)^{2}}{\langle \Delta^{2} x \rangle_{*,t}} 
\end{split}
\end{equation}
This is an instantaneous nonequilibrium uncertainty relation for a process with Gaussian distribution and following a GLE with memory.
Of course, this formula works also for Markov dynamics~\cite{ito18arx}.
  By instantaneous we mean that both the observable $\O$ and the cost function on the right hand side are quantities that depend only on the (PDF of the) position at time $t$.

The cost function of (\ref{eq:main1}) can be related to entropy production rates in some cases discussed in the following section.
We will use expressions from~\cite{OurGLE} for the entropy production rate of the system ($\langle \sigma_{\sys} \rangle_{t_m,t}$), of the environment ($\langle \sigma_{\med} \rangle_{t,t_m}$), and the total one ($\langle \sigma_{\tot} \rangle_{t_m,t}$), all valid for systems described by a Gaussian PDF
\begin{align}\label{entsys} 
  \langle \sigma_{\sys} \rangle_{t_m,t}
  &= \frac{\partial_t \langle \Delta^{2} x \rangle_{t_m,t}}{2\langle \Delta^{2} x \rangle_{t_m,t}}\\
  \label{entmed}
  \langle \sigma_{\med} \rangle_{t_m,t}
  &= \beta \fric(t-t_m) \langle v\rangle_{t_m,t} \langle v_{\text{ret}}\rangle_{t_m,t} -\frac{\beta \SC}{2} \partial_{t} \langle \Delta^{2} x \rangle_{t_m,t}  \\
  \label{enttot}
  \langle \sigma_{\tot} \rangle_{t_m,t} &
  = \langle \sigma_{\sys} \rangle_{t_m,t}+\langle \sigma_{\med} \rangle_{t_m,t}
\end{align}
where 
\begin{align}
\label{v_ret}
\fric(t) & \equiv \int_{0}^{t}\Gamma(t')\mathrm{d}t'\,,\\ \fric & \equiv \displaystyle \lim_{t\to\infty} \fric(t) = \int_{0}^{\infty}\Gamma(t')\mathrm{d}t' < \infty
    \label{gammahat}\\
\langle v_{\text{ret}} \rangle_{t_m,t} &\equiv \frac{1}{\fric(t-t_m)} \int^{t-t_m}_{0} \langle v\rangle_{t_m,t-t'} \Gamma(t')\mathrm{d}t'
\end{align}
are respectively the time dependent friction coefficient, its large time limit and a so called \textit{retarded velocity} (see~\cite{OurGLE}).

\subsection{Particle confined by a harmonic trap}

For an active harmonic trap, we focus on two interesting regimes where the variance of the position \eqref{eq7} is already at its constant asymptotic value,
\begin{align} \label{eqn8a}
\langle \Delta^{2} x \rangle_{t} = \frac{k_{B}T}{\SC}
\end{align}
Indeed, being $\SC$ not modulated and the force $f(t)$ space-independent, the variance of the position cannot be modified.

In the first case, this {\em steady state} for the variance is achieved by starting from $t_{m}\to -\infty$. To justify terming  steady state such regime, we anticipate that we will illustrate it for a particle that is being dragged since a long time by a trap moving at constant velocity. However, the results below hold also for a more complex scenario with general $\lambda(t)$ and $f(t)$.

We show in \ref{sec:limits} that $\lim \limits_{t\to \infty} \chi_{x}(t)=0$ and $\lim \limits_{t\to \infty} \chi(t)=1/\SC$, so that \eqref{eq6} becomes
\begin{align} \label{eqn8}
  \langle x \rangle^{-\infty}_{t} = \int^{t}_{-\infty} \chi_{x} (t-t') F(t') \mathrm{d}t'.
\end{align}
Here the notation $\langle \ldots \rangle^{-\infty}_{t}$ denotes an average obtained for $t_{m}\to -\infty$.
The asymptotic decay of the position susceptibility $\lim \limits_{t\to \infty} \chi_{x}(t)=0$ is expected in a constrained system (this will not be the case for $\SC=0$).

In the second case, equipartition as in (\ref{eqn8a}) holds because we start at $t_m=0$ from an equilibrium distribution under the potential $\frac \SC 2 x^2$.
This implies that $\langle v_{0} \rangle = \langle x_{0} \rangle = 0$, $\langle \Delta^{2}v_{0} \rangle = k_{B}T/m$, $\langle \Delta^{2}x_{0} \rangle = k_{B}T/\SC$ and $\mathrm{Cov}(x_{t_0},v_{t_0})=0$, so that
\begin{align} \label{eqn9}
  \langle x \rangle^{eq}_{t} = \int^{t}_{0} \chi_{x} (t-t') F(t') \mathrm{d}t'.
\end{align}

For these cases of confined particle, the estimates for the average and variance of the position lead to the uncertainty relation
  \begin{equation} \label{eq:main2}
    \mathrm{g}^{\trap}_{\O,t_m}(t) \le
    \mathcal{C}_{t_m}^{\trap}(t)
  \end{equation}
with
  \begin{equation}\label{list1}
  \begin{split}
  & \hspace{3.3cm} \mathrm{g}^{\trap}_{\O,t_m}(t) \equiv \frac{\big \langle \dot{\O} \big \rangle_{t_m,t}^{2}}{\big \langle \Delta^{2} \O \big \rangle_{t_m,t}} \\
  & \mathcal{C}_{t_m}^{\trap}(t) \equiv \frac{\SC}{k_{B} T} ( \partial_t\langle x \rangle_{t_m,t})^{2} = \frac{\SC}{k_{B} T} 
  \left(\chi_{x}(0)F(t)+\int_{t_m}^{t}\chi_{v} (t-t') F(t')\mathrm{d}t' \right)^{2} 
  \end{split}
  \end{equation}
  where $t_m$ is either $0$ or $-\infty$, $\mathcal{C}^{\trap}_{t_m}(t)$ is the cost function and $\mathrm{g}^{\trap}_{\O,t_m}(t)$ is essentially a (squared) signal-to-noise ratio (SNR), a quantity that encodes the precision associated to the observable $R(x)$. Moreover, note that $\chi_{x}(0)=0$ for underdamped dynamics and that the bound \eqref{eq:main2} implies that, for this particular system, the observable with the largest SNR is the position $x$ itself.
  
We have previously shown \cite{OurGLE} that, if the memory kernel is integrable as in \eqref{gammahat},
 for large observation times the entropy production rate of the system becomes
  \begin{equation} \label{equaz24}
   \lim_{\substack{t\to \infty }} \langle \sigma_{\tot} \rangle_{t_m,t}= \lim_{\substack{t\to \infty }} \langle \sigma_{\med} \rangle_{t_m,t}= \frac{\fric}{k_{B}T} \left( \displaystyle\lim_{t\to \infty}\langle v\rangle_{t_m,t}\right)^{2}
   \end{equation}
Hence, for this limit equation \eqref{eq:main2} can be rewritten as
\begin{equation}\label{EntrBound1}
  \lim_{t\to \infty } \mathrm{g}^{\trap}_{\O,t_m}(t) \le
  \frac{\SC}{\fric} \langle \sigma_{\tot} \rangle_{t_m,t} 
\end{equation}
meaning that for long times, instantaneous observables have a SNR $\mathrm{g}^{\trap}_{\O,t_m}(t)$ bounded from above by the mean total entropy production rate times a ratio of the trap strength by the low-frequency damping coefficient.

A particularly interesting regime can be achieved by choosing $\lambda(t)=vt$ and sending $t_m \to - \infty$. Again in \cite{OurGLE}, we show that this can be considered as a steady state for which
\begin{align}\label{Stationary}
    \langle x \rangle^{ss}_{t}=vt-\frac{\fric v}{\SC}\,,
    &&
    \langle v \rangle^{ss}_{t} = \langle v_{\text{ret}} \rangle^{ss}_{t}=v \,,
    && 
    \langle \sigma_{\text{tot}}\rangle_{t}^{ss} = \fric v^2 t \,.
\end{align}
The stochastic inequality hence becomes a TUR for all times
\begin{equation}\label{EntrBoundStaz}
\mathrm{g}^{ss,\trap}_{\O}(t) \le
  \frac{\SC}{\fric} \langle \sigma_{\tot} \rangle^{ss}_{t} 
\end{equation}
Finally, in the Markovian (mk) case, inequality \eqref{EntrBound1} is valid for every $t_m$ and $t$. Indeed in this case 
\begin{align}
  \Gamma^{\text{mk}}(t) = 2\gamma_{0}\delta(t)\,, 
  &&
  \fric^{\text{mk}}(t) = \int_{0}^{t}\Gamma^{\text{mk}}(t)\mathrm{d}t'= \gamma_{0}
  \,,
  &&
  \langle v_{\text{ret}} \rangle_{t_m,t}^{\text{mk}} = \langle v \rangle^{\text{mk}}_{t_m,t}
\end{align}
and from equation \eqref{enttot} we get 
\begin{equation}
    \frac{\SC}{\fric} \langle \sigma \rangle^{\text{mk}}_{t_m,t}  =   \frac{\SC}{k_{B} T} ( \langle v \rangle^{\text{mk}}_{t_m,t})^{2} = \mathcal{C}_{t_m}^{\text{mk},\trap}(t) \end{equation}
in which, again, the cost function is always proportional to the entropy production rate.

\subsection{Particle not confined}

When no confinement is present ($\SC = 0$), the only way to drive our system out of equilibrium is through the space-independent force $f(t)$.
We again analyse two situations.

First, we consider an initial distribution that can be factorised as $P(x_{t_0},v_{t_0},t_0)=\delta(x-x_0)P^{eq}(v_{0})$ (a Dirac delta is a limit of a Gaussian) and we find that
\begin{align}\label{eqn10}
 \langle x \rangle^{dd}_{x_0,t} & =
  x_{0} +\int^{t}_{0} \chi_{x} (t-t') F(t') \mathrm{d}t'\,, &&
  \langle \Delta^{2} x \rangle^{dd}_{t} = 2 k_{B} T \chi(t)\,,
\end{align}
where we used that $\langle \Delta^2 x_0 \rangle = 0$ and $\mathrm{Cov}(x_{0},v_{0})=0$. Experimentally, this can obtained by selecting any occurrence where $x(t) =x_{0}$ and use it as an initial point for the future dynamics. Note that this kind of initial distribution could have been also used for the constrained case but we simply chose not to consider it.

Otherwise, we can prepare the system in an initial equilibrium distribution with an optical trap, say with stiffness $\SC'$, and switch it off when the external force is turned on. This would correspond to  $\langle v_{0} \rangle = \langle x_{0} \rangle = 0$, $\langle \Delta^{2}v_{0} \rangle = k_{B}T/m$, $\langle \Delta^{2}x_{0} \rangle = k_{B}T/\SC'$ and $\mathrm{Cov}(x_{t_0},v_{t_0})=0$, so that 
\begin{align}\label{eqz10}
 \langle x \rangle^{\SC'}_{x_0,t} & =
 \int^{t}_{0} \chi_{x} (t-t') F(t') \mathrm{d}t', && 
 \langle \Delta^{2} x \rangle^{\SC'}_{t} = k_{B} T ( 2 \chi(t) + 1/\SC' )
\end{align}
We underline that, in all the cases shown above, the average and the variance fully characterize the PDF of the position, as its Gaussian character is preserved by construction.  

By using equations \eqref{eqn10} and \eqref{eqz10} we get 
\begin{equation} \label{eq:main3}
  \mathrm{g}^{\diff}_{\O}(t) \le
  \mathcal{C}^{\diff}(t)
\end{equation}
with
\begin{equation}\label{list2}
  \begin{split}
 &\hspace{3cm} \mathrm{g}^{\diff}_{\O}(t) \equiv \frac{\big \langle \dot{\O} \big \rangle_{t}^{2}}{\big \langle \Delta^{2} \O \big \rangle_{t}/t} \\
  \frac{\mathcal{C}^{\diff}(t)}{t} \equiv & \frac{1}{2} \left(\frac{\partial_t\langle \Delta^{2} x \rangle_{t}}{\langle \Delta^{2} x \rangle_{t}}\right)^2 + \frac{\left( \partial_t\langle x \rangle_{t} \right)^{2}}{\langle \Delta^{2} x \rangle_{t}} =\\
  = & \left(\frac{\chi_{x} (t)}{ 2 \chi(t) + \langle \Delta x_{0} \rangle }\right)^2 + \frac{\left(\chi_{x}(0)F(t)+\int_{0}^{t}\chi_{v} (t-t') F(t') \mathrm{d}t' \right)^{2}}{k_{B} T ( 2 \chi(t) + \langle \Delta x_{0} \rangle)} 
  \end{split}
  \end{equation}
  where $\langle \Delta x_{0} \rangle=0$ for an initial distribution that is factorised into a Dirac delta for the initial position or $\langle \Delta x_{0} \rangle=k_{B}T/\SC'$ if the system is at equilibrium at time $t=0$ in an harmonic trap of stiffness $\SC'$. 
  
  Note that the SNR $\mathrm{g}^{\diff}_{\O}(t)$ and the cost function $\mathcal{C}^{\diff}(t)$ for this diffusive case are defined differently from \eqref{list1}. Now there is an additional factor $t$ to let these quantities converge to a constant value at large observation times. To analyse this regime, we use the limits of susceptibilities calculated in \ref{sec:limits}, in particular the fact that $\displaystyle\lim_{t\to\infty}\chi(t)= t/\fric$, along with the expressions for the entropy production rates \eqref{entsys} and \eqref{entmed}. In doing this, it is straightforward to see that 
\begin{equation}
  \displaystyle\lim_{t \to \infty} \mathcal{C}^{\diff}(t) =  \frac{\partial_t\langle \Delta^{2} x \rangle_{t}}{2\langle \Delta^{2} x \rangle_{t}} + \frac{\fric}{2} \langle v \rangle_{t}^{2} = \frac{1}{2}(\langle \sigma_{\tot} \rangle_{t} +\langle \sigma_{\sys} \rangle_{t})
\end{equation}
where we also used equation \eqref{equaz24}. Hence for $t \to \infty$ the bound \eqref{eq:main3} becomes 
\begin{equation}\label{EntrBound2}
    \lim_{t\to\infty}\mathrm{g}^{\diff}_{\O}(t) \le
    \frac{1}{2}(\langle \sigma_{\tot} \rangle_{t} +\langle \sigma_{\sys} \rangle_{t})
\end{equation}
Note that the bound above becomes valid for all times in the special case of Markovian dynamics in the overdamped limit and starting from an initial distribution that is a Dirac delta.

\section{Applications}
\label{sec:appl}

We discuss some regimes for which it is possible to derive explicit analytical expressions for the SNRs and for the cost function or the entropy production rates.

\subsection{Exponential memory kernel with confinement}

\begin{figure}[tb]
\begin{center}
\includegraphics[width=0.94\textwidth, angle=0]{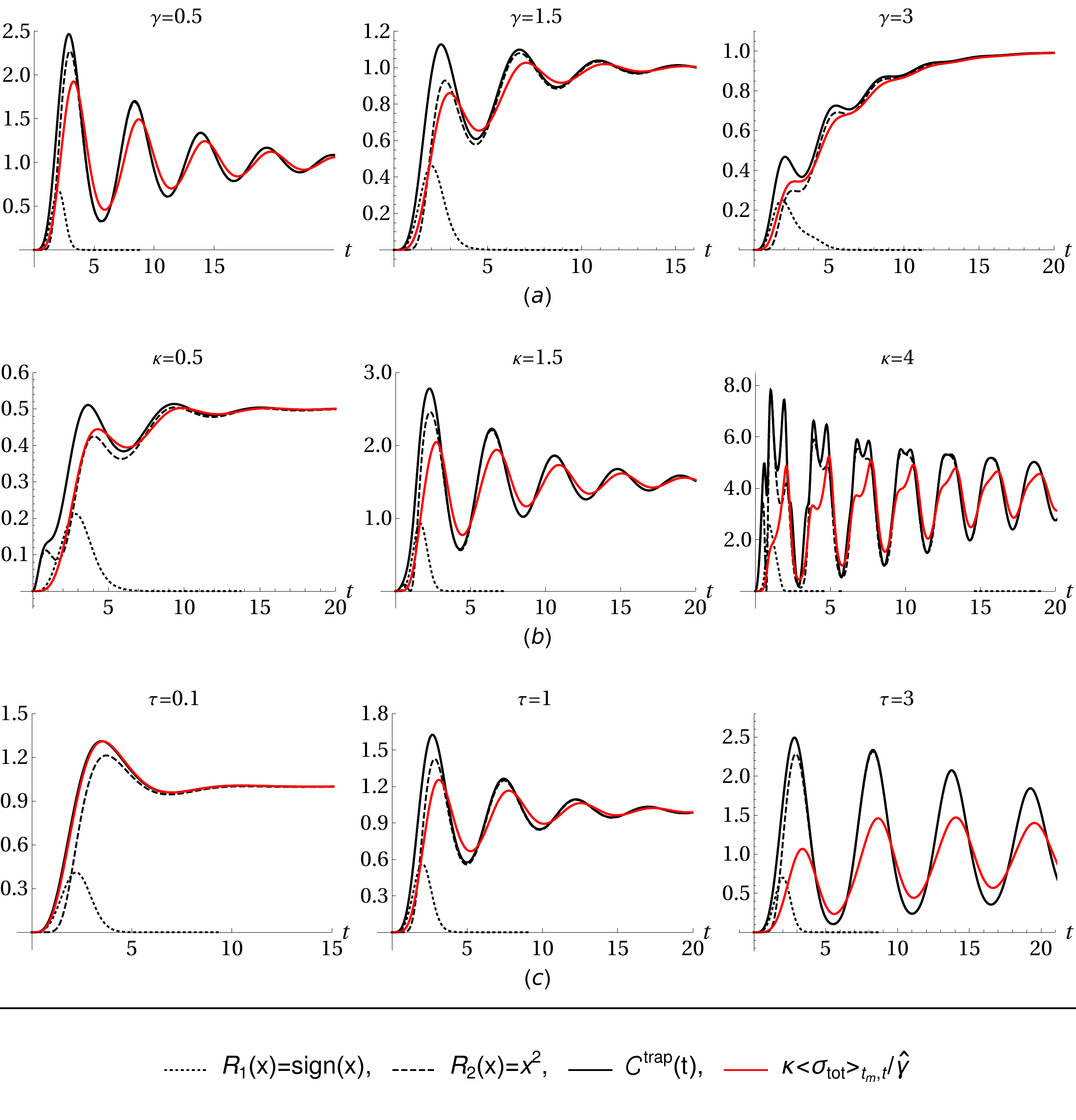}
\end{center}
\caption{For a GLE with pure exponential kernel ($\gamma_{0}=0$), and starting from an equilibrium distribution:
  SNRs $g^{\trap}_{R} = \langle \dot{R} \rangle^{2}_{t}/\langle \Delta^{2}R \rangle_t$ (see legend), cost function and entropic bound (red line).
With reference to parameters $m=1$, $v=1$, $\SC =1$, $\gamma=1$ and $\tau=1$, 
in (a) we vary $\gamma$ (quantities exhibit oscillations that become less pronounced as $\gamma$ grows), in (b) we vary $\SC$ (oscillations become stronger and more persistent in time as $\SC$ becomes larger and the limit of the cost functions as well as the entropic bounds grow linearly as the value of trap stiffness becomes larger), and in (c) we vary $\tau$  (note that for $\tau=0.1$, i.e.~at quasi-Markovianity, red and black continuous lines nearly coincide: this reflects the fact that for Markovian dynamics the cost function becomes proportional to the rate of entropy production).
In all cases, for large times the cost function converges to the entropy production rate. However, we have visualized that in general the entropy production rate is not a bound for the SNRs.
}
\label{Fig1}
\end{figure}

\begin{figure}[tb]
\begin{center}
\includegraphics[width=0.94\textwidth, angle=0]{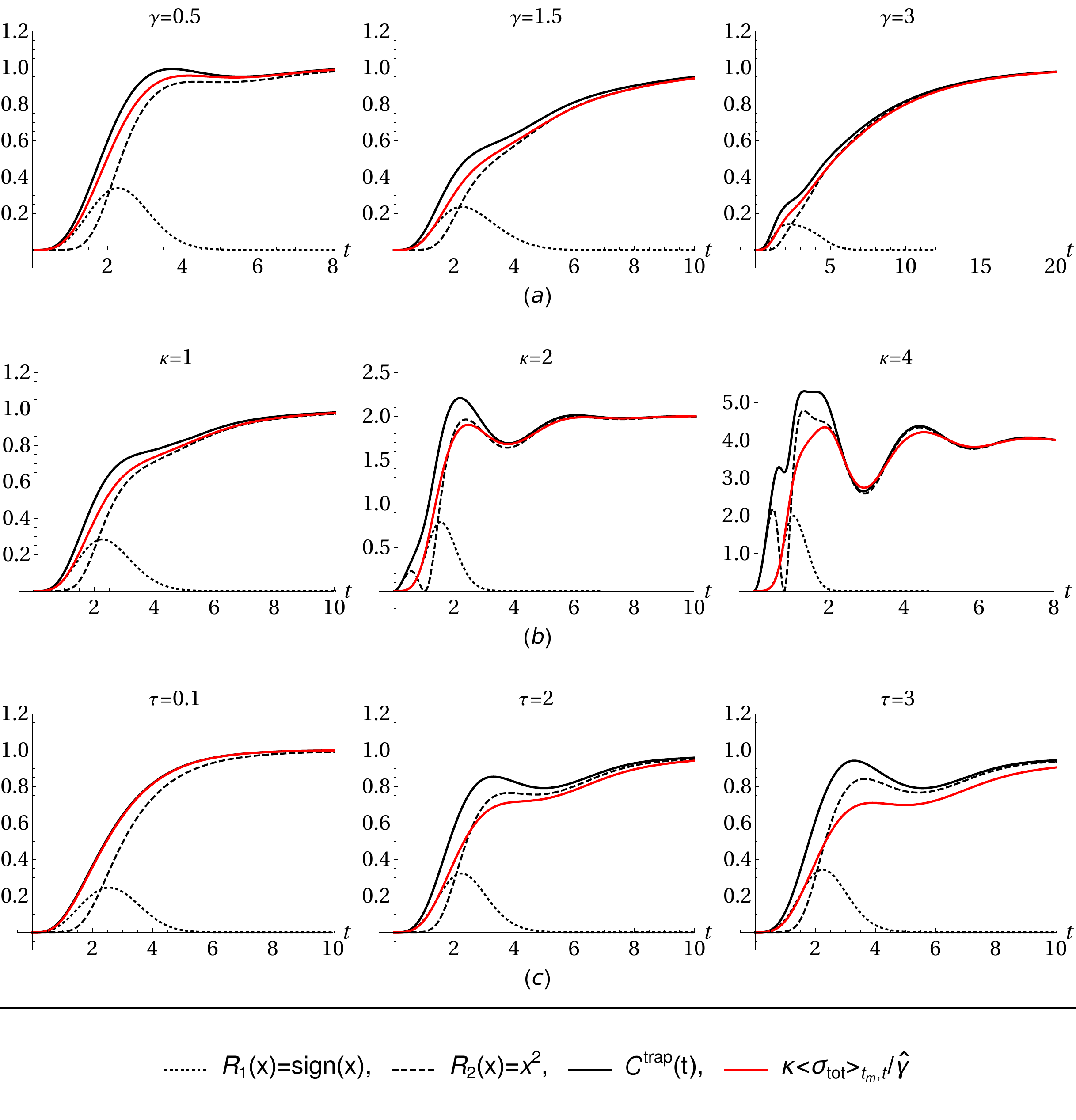}
\end{center}
\caption{Similar to Figure~\ref{Fig1} but for a complete exponential memory kernel with Markovian contribution $\gamma_{0}=0.5$ that smothers the oscillations.
}
\label{Fig2}
\end{figure}

For our example, we focus on a simple memory kernel with one exponential component, with GLE 
\begin{equation} 
 m\ddot x (t) = - \gamma_{0} \dot x (t)- \frac{\gamma}{\tau}\int_{t_{m}}^{t} \e^{- (t-t')/\tau} \dot x(t') \mathrm{d}t'- \SC \left[ x (t) - vt\right] + \eta (t)
\end{equation} 
where we set $\lambda(t)=vt$ and $f(t)=0$. As already hinted in the previous section, with this linear dragging protocol a steady state for $t_m \to -\infty$.

Here we analyse the the bound \eqref{eq:main2} for two different observables, i.e.~$\O_{1}(x) = {\rm sign}(x)$ and $\O_{2}(x)=x^2$, starting from equilibrium or from a stationary state. In the latter case, the bound becomes a full-fledged entropic bound \eqref{EntrBound1}.

As a first standard example for viscoelastic fluids, we analyze the case of an exponential memory kernel,
\begin{align}
  \Gamma^{\exp} (t) = \gamma_{0}\delta(t)+ \sum_{i=1}\frac{\gamma_{i}}{\tau_{i}}\e^{-t/\tau_{i}}\,, &&
  \hat{\Gamma}^{\exp} (k) = \gamma_{0}+ \sum_{i=1}\frac{\gamma_{i}}{1+k\tau_{i}}\,,
\end{align}
with
\begin{align}
  \fric = \int_{0}^{\infty}\Gamma(t')\mathrm{d}t'= \sum_{i=0} \gamma_{i}\,, && 0\le\fric<\infty \,.
\end{align}
This is an important example, as a finite sum of suitably sized exponential terms can approximate, up to a finite time scale, every memory kernel even if $\fric$ does not converge, see \cite{ViscGLE_Goy} for details.

\begin{figure}[tb]
\begin{center}
\includegraphics[width=0.94\textwidth, angle=0]{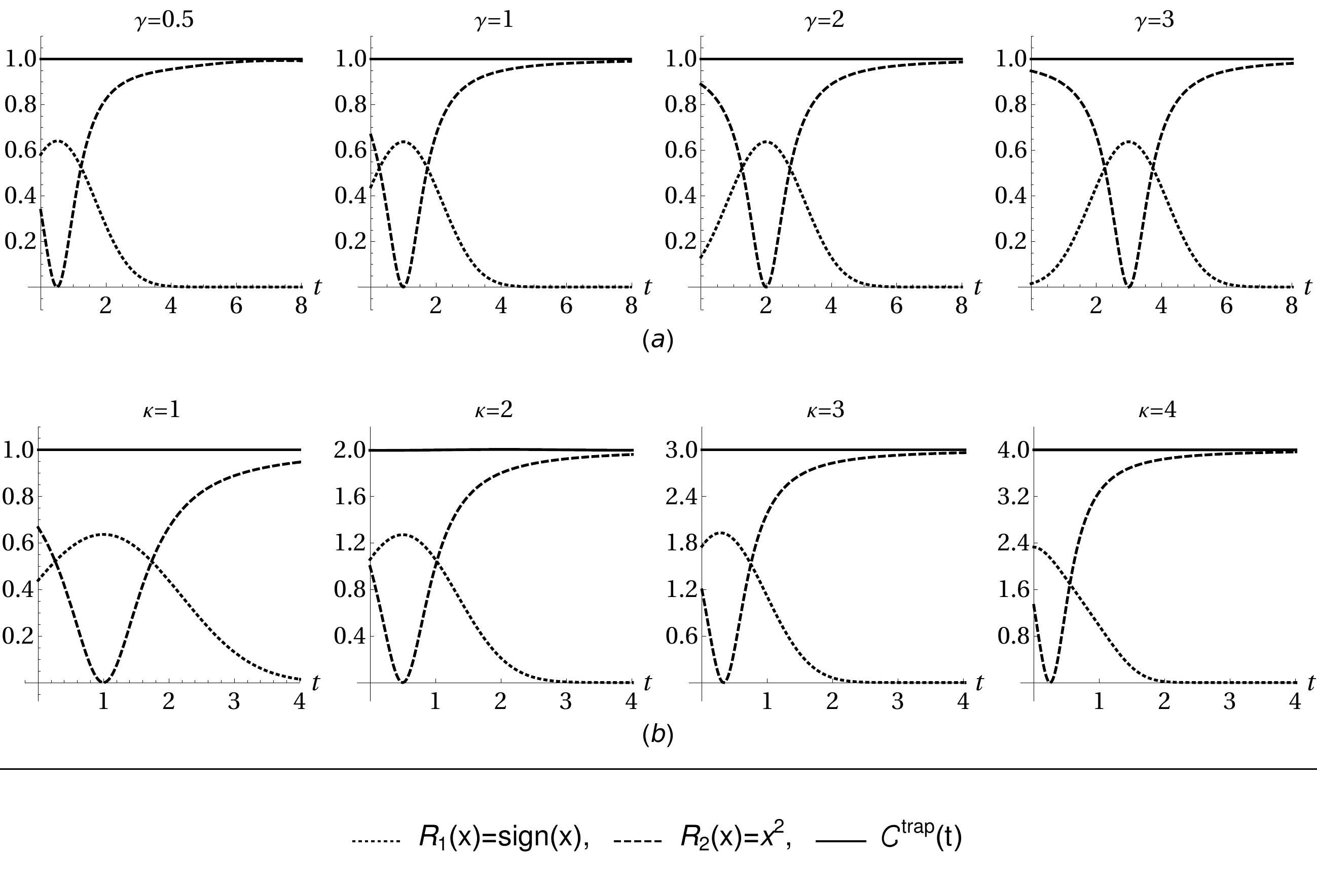}
\end{center}
\caption{As in Figure \ref{Fig1} but for a trap dragging the particle with constant velocity (the ``steady state'') and passing with its minimum $\lambda(t)= v t$ at $\lambda(0)=0$. In this case the cost function $\mathcal{C}^{\trap,\text{ss}}(t)
 =\frac{\SC}{\fric} \langle \sigma_{\tot} \rangle_{t}^{\text{ss}}= \SC v^2 $ matches the entropic bound, proportional to the constant entropy production rate (see equation \eqref{Stationary}).
  Variations of memory characteristic time are not considered as their effects are not present as $t_m \to - \infty$. 
  (a) A larger value of $\gamma$ corresponds to a shift of the minimum and maximum of the two SNRs towards larger observation times. The cost function remains unaffected by variations of $\gamma$.
  (b) As the trap stiffness grows, so does the cost function proportionally, while the minimum and maximum of the SNRs move towards smaller observation times.
  }
\label{Fig3}
\end{figure}

For a memory kernel that is purely exponential, i.e.~when $\gamma_{0}=0$, we note that the SNRs as well as the bounds exhibit strong oscillations when starting from an equilibrium distribution, depending of course on the values of the parameters (Figure~\ref{Fig1}). When $\gamma_{0} \neq 0$ instead, these oscillations are smothered (Figure~\ref{Fig2}). No significant difference is seen instead if we start from a stationary state (thus we show only the case $\gamma_0=0$ in Figure~\ref{Fig3}), in fact if $t_m \to - \infty$ the memory effects are lost and the dynamics only depends on the limit of the time dependent friction coefficient $\fric$, see equation \eqref{Stationary}. In other words it is not possible anymore to distinguish the effects of the exponential part of the memory kernel from the Markovian one.

\subsection{Exponential memory kernel without confinement}

\begin{figure}[tb]
\begin{center}
\includegraphics[width=0.94\textwidth, angle=0]{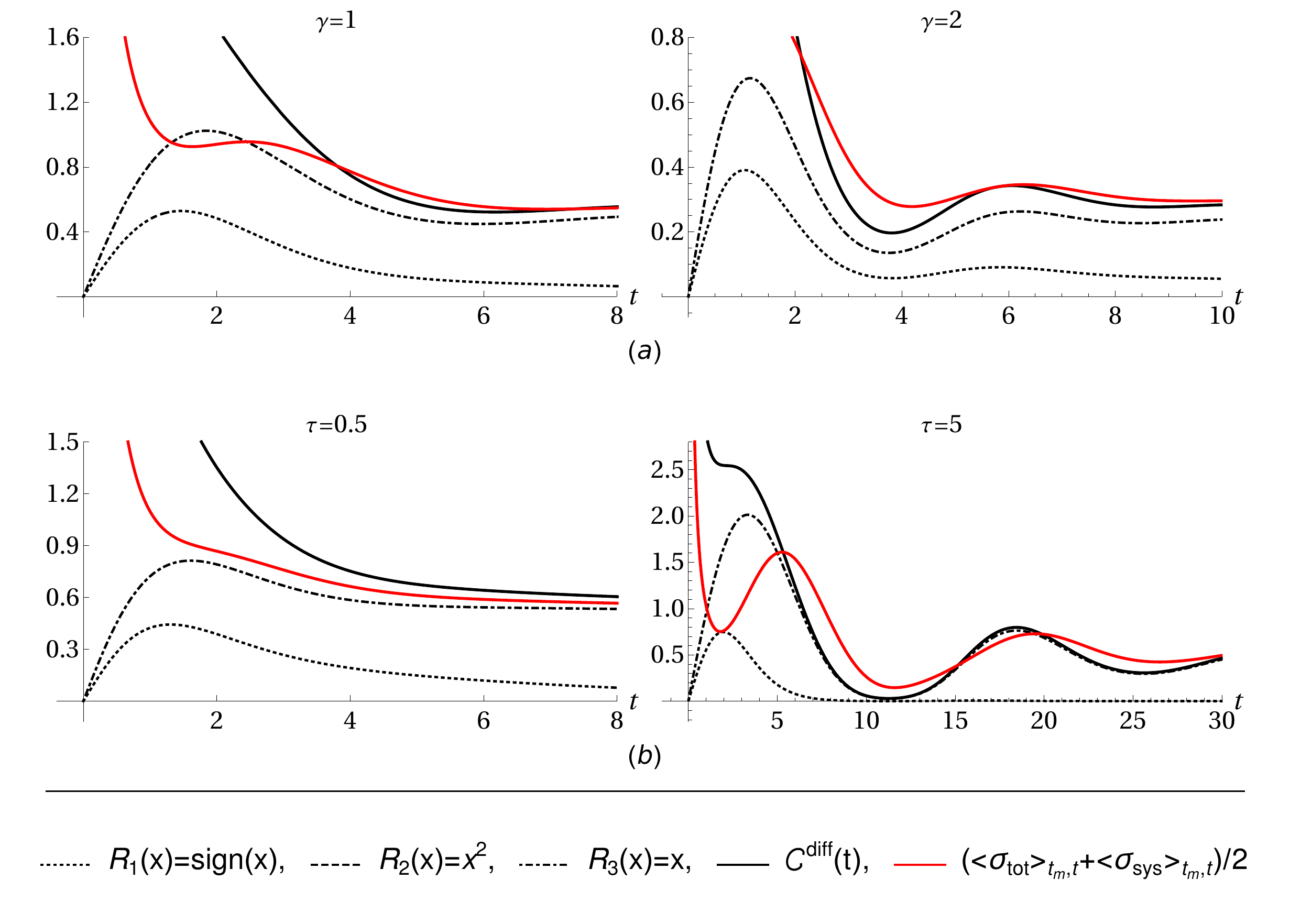}
\end{center}
\caption{For free diffusion ($\SC=0$) under a constant force $f=1$ from an initial distribution $P(x_{t_0},v_{t_0},t_0)=\delta(x-x_0)P^{eq}(v_{0})$: SNRs $g^{\diff}_{R} = t\langle \dot{R} \rangle^{2}_{t}/\langle \Delta^{2}R \rangle_t$ (see legend), cost function and entropic bound (red line) for pure exponential memory kernel ($\gamma_{0}=0$). As in previous figures, $m=1$,  $\gamma=1$ and $\tau=1$.
  In row (a) we vary $\gamma$.
  Differently from the bounded case (see Figure \ref{Fig1}) oscillations become stronger in amplitude as $\gamma$ increases while again the limit to which the cost functions and the entropy production rates does not change with $\gamma$. In (b) instead we vary $\tau$. As before, the long time limit of the cost function is approached also by the corresponding entropic bound, while oscillations increase as the memory characteristic time gets larger.
  The bound is very quickly saturated for the observable $\O_{2}(x)=x^2$, hence its SNR is not visible in the panels.}
\label{Fig.diff.del.0}
\end{figure}

\begin{figure}[tb]
\begin{center}
\includegraphics[width=0.94\textwidth, angle=0]{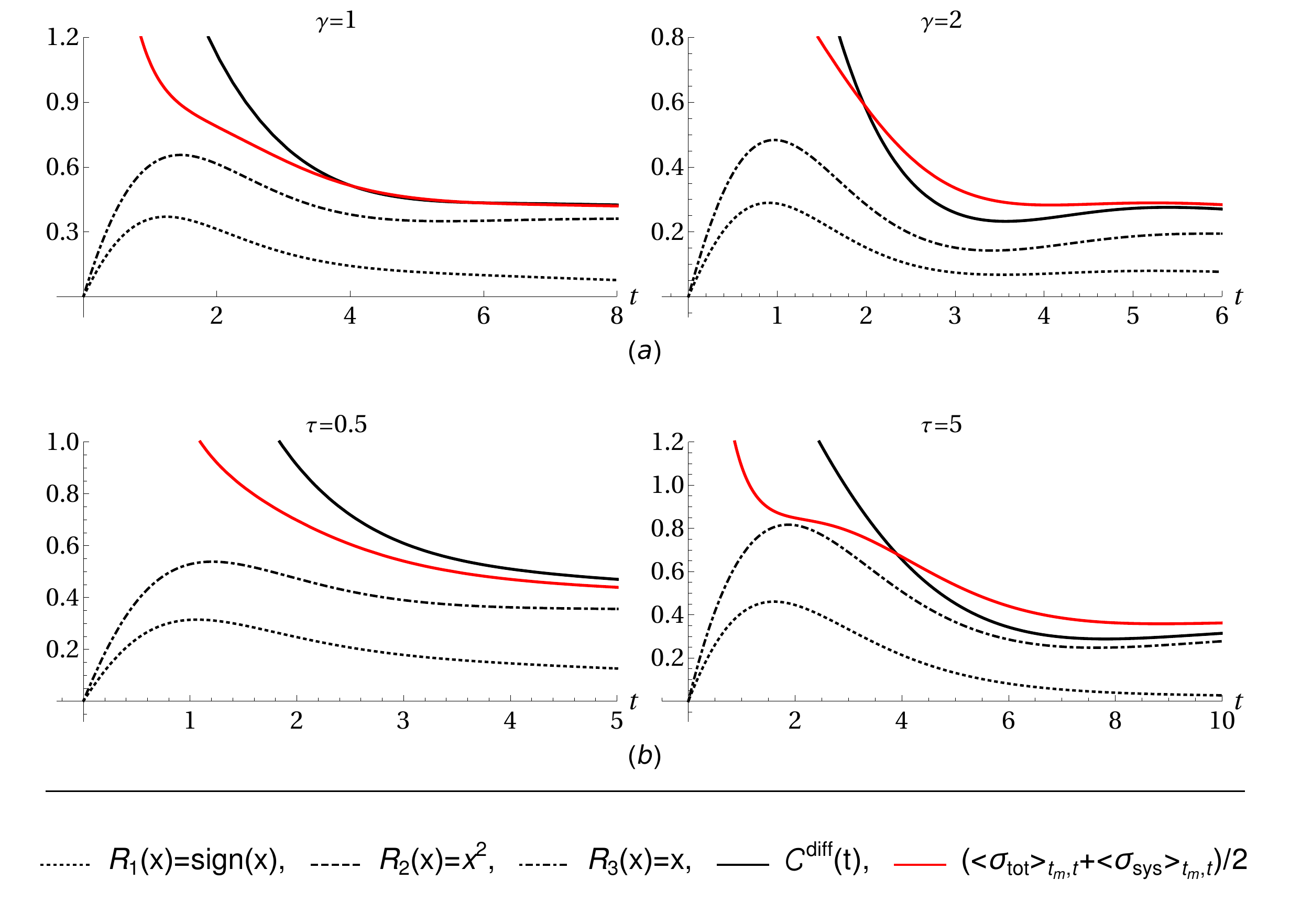}
\end{center}
\caption{As in Figure~\ref{Fig.diff.del.0} but for complete exponential memory kernel ($\gamma_{0}=0.5$, damping the oscillations) with initial distribution $P(x_{t_0},v_{t_0},t_0)=\delta(x-x_0)P^{eq}(v_{0})$.
}
\label{Fig.diff.del.1}
\end{figure}

Figure \ref{Fig.diff.del.0} and Figure~\ref{Fig.diff.del.1} show the case of an initial distribution that is a Dirac delta for the starting position, which implies $\langle \Delta^2 x_{0} \rangle=0$. For small times this causes a divergence of the cost function $\mathcal{C}^{\diff}(t)$ $t$ due to its term $(\partial{t}\langle \Delta^2 x \rangle_t / \langle \Delta^2 x \rangle_t)^{2}$. While in this regime the bound becomes loose for $\O_{1}(x)$ and $\O_{3}(x)$ it is immediately saturated for $\O_{2}(x)$.

We analyse diffusion dynamics ($\SC=0$) of the bead subject to an external force $f(t)=f$ that is constant both in space and time.
The variance of the position grows in time, hence there exists no stationary distribution. Also the average position grows linearly in time due to $f$.

As in the previous subsection, for simplicity the associated GLE contains a single exponential with characteristic time $\tau$,
\begin{equation} 
 m\ddot x (t) = - \gamma_{0} \dot x (t)- \frac{\gamma}{\tau}\int_{0}^{t} \e^{- (t-t')/\tau} \dot x(t') \mathrm{d}t' +f + \eta (t)
\end{equation} 
We thus discuss the bound \eqref{eq:main3} for the dynamics generated by the above equation, again noting the entropic nature of the bound in the large time limit. We will consider three observables: $\O_{1}(x) = {\rm sign}(x)$, $\O_{2}(x)=x^2$ and $\O_{3}(x)=x$. The latter observable has a non-saturating SNR for this unbound diffusion dynamics.

\begin{figure}[tb]
\begin{center}
\includegraphics[width=0.94\textwidth, angle=0]{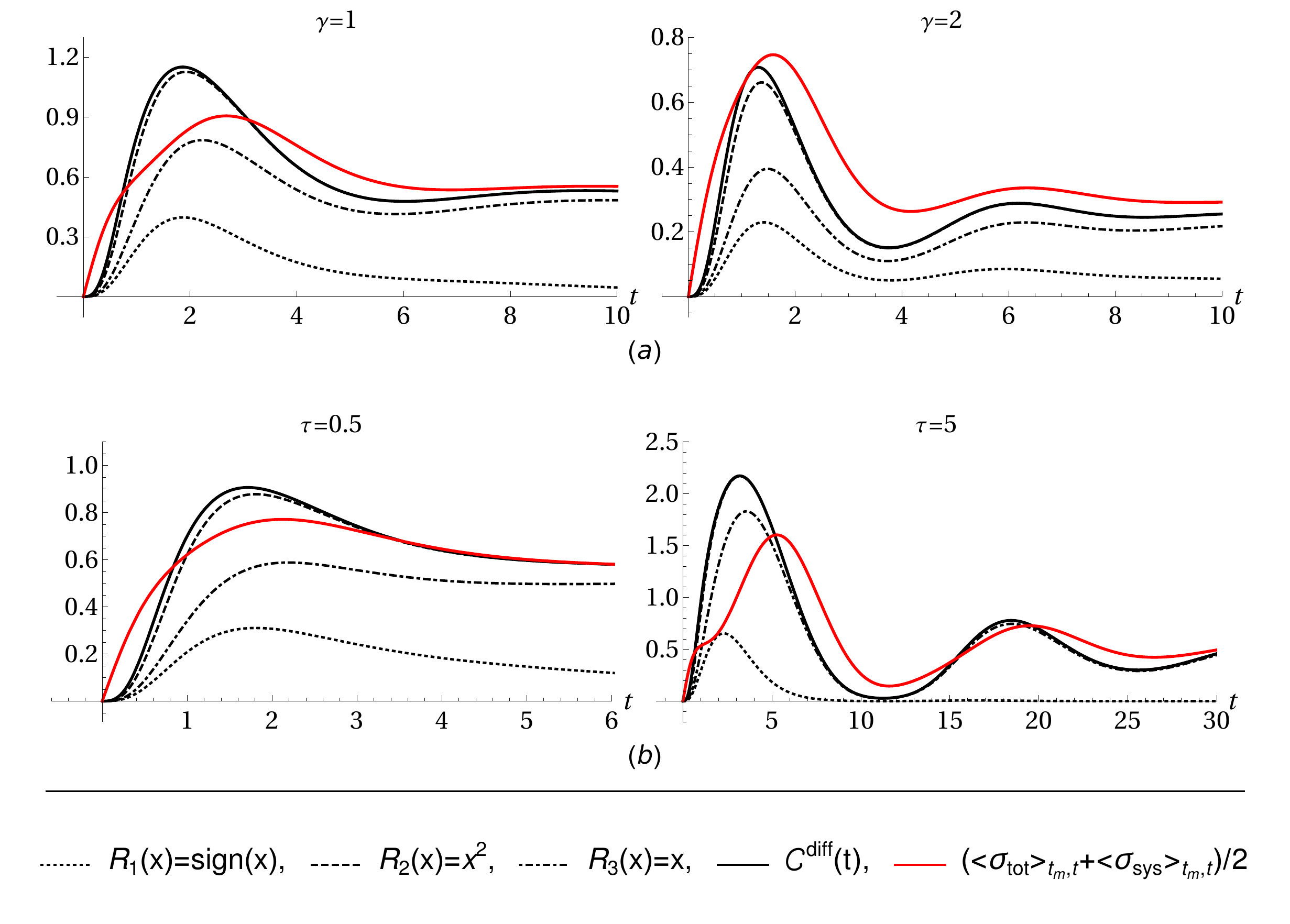}
\end{center}
\caption{SNRs $g^{\diff}_{R} = t\langle \dot{R} \rangle^{2}_{t}/\langle \Delta^{2}R \rangle_t$, cost function and entropic bound for pure exponential memory kernel ($\gamma_{0}=0$) and for initial distribution $P(x_{t_0},v_{t_0},t_0)=P^{\SC',eq}(v_{0})P^{eq}(v_{0})$ with $\langle \Delta^{2}x_{0}\rangle = k_{B}T/\SC'$. In row (a) we chose $m=1$, $f=1$ and $\tau=1$. A finite initial variance of the position avoids a divergence of the cost functions. Differently from the bounded case (see Figure \ref{Fig1}) oscillations become stronger in amplitude as $\gamma$ increases while again the limit to which the cost functions and the entropy production rates does not change with $\gamma$. For (b) instead we have $\gamma=1$, $m=1$ and $f=1$. Like before the large time limit of the cost function and rates is the same for both values of $\tau$ while oscillations increase as the memory characteristic time grows larger. Again the bound is very quickly saturated for $\O_{2}(x)$.}
\label{Fig.diff.eq.0}
\end{figure}

However, if the dynamics starts from an equilibrium condition in an optical trap of stiffness $\SC'$ (implying $\langle \Delta^{2}x_{0}\rangle = k_{B}T/\SC'$) no divergences occur and the bound becomes tighter for all observables. This can be all seen in Figure~\ref{Fig.diff.eq.0}, for $\gamma_0=0$.
The case $\gamma_0>0$ yields similar plots with less oscillations.

To summarize, the entropic bound is violated for finite times and is only valid asymptotically.
It is perhaps surprising that the observable which goes closer to saturate the inequality is $\O_{2}=x^2$ instead of $\O_{3}=x$, which fully saturates the bound for trapped dynamics.

\section{Conclusions}
\label{sec:concl}

Considering a system as optical tweezers dragging a microbead in a complex fluid,
we have derived a nonequilibrium inequality \eqref{eq:main1} for Langevin equations with memory kernel, for the cases in which the position evolves distributed as a Gaussian. This inequality covers also diffusion not bounded by a harmonic trap but driven by a homogeneous time-dependent field.

The inequality \eqref{eq:main1} quantifies how the signal-to-noise ratio of observables is bounded by a cost function. By focusing on instantaneous quantities as the bead position, it is in line with a recent TUR for Markovian dynamics~\cite{koy20} and embodies a previous Markovian version~\cite{ito18arx}. An approach based on instantaneous quantities is a viable option for dealing with non-Markovian systems, which have more complicated path weights than those of Markovian systems. 

The cost function in the inequality \eqref{eq:main1} in general is not the entropy production, but it can become proportional to the entropy production rate in some limits. For a particle confined by a harmonic trap, \eqref{eq:main1} can be cast as \eqref{eq:main2}, which becomes the TUR \eqref{EntrBound1} in the limit of large observation times (again, this TUR contains the instantaneous entropy production rate, at variance with TURs in the literature). Moreover, for a particle dragged at constant velocity in a complex fluid by moving optical tweezers, the TUR \eqref{EntrBoundStaz} holds for all times if the dragging has been performed since a long time before the beginning of measurements.

For particles not constrained by optical tweezers, but eventually subject to a global force $f(t)$, the inequality reduces to \eqref{eq:main3}. This may also become an instantaneous TUR with cost function proportional to some entropy production rates, see \eqref{EntrBound2}. For instance, for integrable memory kernels, at long times the effects of the memory are lost and essentially the system behaves as a Markovian one. 

Markovian Langevin dynamics is a particular subclass of what we have described. In all cases we have analysed, a Markovian dynamics may lead more easily to a thermodynamic interpretation in which the entropy production rate is the function bounding the signal-to-noise ratio (besides constant prefactors). Indeed, we recover it for a particle dragged but starting from equilibrium, and also for a free overdamped particle starting from a given position (Dirac delta distribution initially). In the latter case the susceptibility is equal to  $\chi(t)=2 D t$, with diffusion constant $D$. Thus $\chi_x(t) = \partial_t \chi(t) = 2 D$
and the ratio $(t\chi_x(t)) / \chi(t) = 1$, so that from \eqref{EntrBound2} we can infer the validity of  the instantaneous TUR \eqref{EntrBound2} for all times.
A realisation of this scenario could be a charged particle driven by a homogeneous time-varying electric field in a fluid without memory \eqref{EntrBound2}.

The fact that a nonequilibrium inequality contains a cost function not directly related only to the entropy production is not surprising. As discussed in the introduction, many previous examples show that other nondissipative aspects may be constraining the signal-to-noise ratio of observables, in conjuction with or in alternative to the entropy production.

Our calculations were performed by remaining within the domain of Gaussian statistics. It seems interesting to check if and how one could generalize these results by preserving the scheme of observations that are local in time.

\ack We thank Andreas Dechant, Gianmaria Falasco and Christian Maes for useful discussions.

\appendix

\section{Limits of susceptibilities }\label{sec:limits}
In this section we discuss the limits of the position susceptibility, a very useful quantity that appears throughout the whole article and defined as
\begin{equation}
\hat{\chi}_{x}(k) = [m k^{2}+k\hat{\Gamma}(k)+\SC]^{-1}
\end{equation}
along with the limit of its integral and derivative
\begin{align} 
  \chi(t) \equiv \int_{0}^{t}\chi_{x}(t') \mathrm{d}t'\,, && \chi_{v} (t) \equiv \partial_{t} \chi_{x}(t)\,,
\end{align}
both in the underdamped and overdamped case. To this end, we use the Tauberian theorem for Laplace transforms, which states that, for a given function $g(t)$,
\begin{align}
  \displaystyle\lim_{t\to 0} g(t) = \mathcal{L}^{-1}_{t} \left[\displaystyle\lim_{k\to \infty} \hat{g}(k) \right]\,, && \displaystyle\lim_{t\to \infty} g(t) = \mathcal{L}^{-1}_{t} \left[\displaystyle\lim_{k\to 0} \hat{g}(k) \right] \,.
\end{align}
Furthermore, we focus on the memory kernel used in section \ref{sec:appl}, i.e.
\begin{align}
  \Gamma^{\exp} (t) = \gamma_{0}\delta(t)+ \sum_{i}\frac{\gamma_{i}}{\tau_{i}}\e^{-t/\tau_{i}} \,,
\end{align}
where $\Eul$ is again the Euler gamma. The Laplace transform of this the memory kernel is
\begin{align} \label{lim3}
  \hat{\Gamma}^{\exp} (k) = \gamma_{0}+ \sum_{i}\frac{\gamma_{i}}{1+k\tau_{i}} \,.
\end{align}
We first consider the long time limit of the susceptibilities, distinguishing between the bounded ($\SC \ge 0$) and free ($\SC=0$) case. We have 
\begin{equation} \label{lim6}
\begin{split}
  & \hspace{0.3cm}\displaystyle\lim_{t\to \infty} \chi_{x}(t) = \mathcal{L}^{-1}_{t} \left[\displaystyle\lim_{k\to 0} \frac{1}{m k^{2}+k\hat{\Gamma}(k)+\SC} \right] \approx \mathcal{L}^{-1}_{t} \left[ \frac{1}{\SC} \right] = \frac{\delta(t)}{\SC} \stackrel{t \to \infty}{=} 0 \\
  & \displaystyle\lim_{t\to \infty} \chi(t) = \mathcal{L}^{-1}_{t} \left[\displaystyle\lim_{k\to 0} \frac{1}{k(m k^{2}+k\hat{\Gamma}(k)+\SC)} \right] \approx \mathcal{L}^{-1}_{t} \left[ \frac{1}{k \SC} \right] = \frac{\theta(t)}{\SC} \stackrel{t \to \infty}{=} 1/\SC
\end{split}
\end{equation} 
where we used that $\chi(t)=\int_{0}^{t}\chi_{x}(t')\mathrm{d}t'$.
The free case ($\SC=0$) instead is drastically different, in fact we get that 
\begin{equation} \label{lim7}
\begin{split}
  & \hspace{0.2cm} \displaystyle\lim_{t\to \infty} \chi_{x}(t) = \mathcal{L}^{-1}_{t} \left[\displaystyle\lim_{k\to 0} \frac{1}{m k^{2}+k\hat{\Gamma}(k)} \right] \approx \mathcal{L}^{-1}_{t} \left[ \frac{1}{k \hat{\Gamma}(0)} \right] = 1/\fric \\
  & \displaystyle\lim_{t\to \infty} \chi(t) = \mathcal{L}^{-1}_{t} \left[\displaystyle\lim_{k\to 0}\frac{1}{ k(m k^{2}+k\hat{\Gamma}(k))} \right] \approx \mathcal{L}^{-1}_{t} \left[ \frac{1}{k^2 \hat{\Gamma}(0)} \right] = t/\fric
\end{split}
\end{equation} 
where we noted that $\hat{\Gamma}(0)= \fric = \int_{0}^{\infty}\Gamma(t')\mathrm{d}t'$.
Moreover, note that all this limits do not depend on $m$ and hence they are valid also in the overdamped limit. Things become different in the limit of $t\to 0$.

\subsection{Underdamped}
Applying the Tauberian theorem to the underdamped position susceptibility, we get
\begin{equation}
  \displaystyle\lim_{t\to 0} \chi_{x}(t) = \mathcal{L}^{-1}_{t} \left[\displaystyle\lim_{k\to \infty} [m k^{2}+k\hat{\Gamma}(k)+\SC]^{-1} \right] \approx \mathcal{L}^{-1}_{t} \left[ \frac{1}{m k^{2}} \right] = \frac{t}{m} \stackrel{t \to 0}{=} 0
\end{equation}
where we used that in \eqref{lim3} it holds that
\begin{equation}
\lim_{k\to \infty} \frac{m k^2}{k\hat{\Gamma}(k)} \gg 1 \,.
\end{equation}
As for its integral of course we have that
\begin{equation}
  \displaystyle\lim_{t\to 0} \chi(t) = \displaystyle\lim_{t\to 0} \int_{0}^{t}\chi_{x}(t')\mathrm{d}t' \approx \frac{t^{2}}{2m} \stackrel{t \to 0}{=} 0
\end{equation}
We see that this result does not depend on the form of the kernel. In fact, inertial effects dominate the particle behaviour in the small time limit.

\subsection{Overdamped}

Overdamped dynamics is obtained by performing the massless limit $m \to 0$ so that the Laplace transform of the position susceptibility becomes $\hat{\chi}_{x}^{ov}(k)=[k\hat{\Gamma}(k)+\SC]^{-1}$. We have that
\begin{equation}\label{lim9}
\displaystyle\lim_{t\to 0} \chi^{ov, \exp}_{x}(t) = \mathcal{L}^{-1}_{t} \left[\displaystyle\lim_{k\to \infty} [k\hat{\Gamma}^{\exp}(k)+\SC]^{-1} \right] \approx \mathcal{L}^{-1}_{t} \left[ \frac{1}{k \gamma_{0}} \right] = \frac{1}{\gamma_{0}} \\
\end{equation}
As it can be seen from first line of the last equation, in the overdamped limit it is important that $\Gamma^{\exp}(t)$ has a piece proportional to the Dirac delta,
 as pointed out in \cite{nascimento2019non}.
Finally, for the integral of the susceptibility we get
\begin{equation}
  \displaystyle\lim_{t\to 0} \chi^{ov, \exp}(t) = \mathcal{L}^{-1}_{t} \left[\displaystyle\lim_{k\to \infty} \frac{1}{k(k\hat{\Gamma}^{\exp}(k)+\SC)} \right] \approx \mathcal{L}^{-1}_{t} \left[ \frac{1}{k^2 \gamma_{0}} \right] = \frac{t}{\gamma_{0}}
\end{equation}

Note that in the small time limit the trapping plays no role as all the susceptibilities do not depend on $\SC$.

\section{Kullback-Leibler divergence}\label{sec:AppB}
 
We show how to calculate equation \eqref{eq13}.
We recall that the definition of Kullback-Leibler divergence between two PDFs $P^\alpha$ and $P$ is
\begin{equation}
  \KL( t |*) = \int \mathrm{d} x P^{\alpha} \left( x, t|* \right) \ln \left[ \frac{ P^{\alpha} \left( x, t | *\right)}{ P \left( x, t|* \right)}\right]
\end{equation}
In our case $ P \left( x, t|* \right)$ is assumed to be a Gaussian PDF dependent on some initial conditions $*$ (and so will the corresponding average and variance) and that $P^{\alpha}\left(x,t|*\right)$ is a perturbed Gaussian PDF dependent on a small perturbation parameter $\alpha \approx 0$. 
For two Gaussian distributions 
\begin{align}
P \left( x , t |*\right) &
    = \frac{1}{\sqrt{2 \pi \langle \Delta^{2} x \rangle_{*,t}} } \exp \bigg[ - \frac{\left( x - \langle x \rangle_{*,t} \right)^{2}}{2 \langle \Delta^{2} x \rangle_{*,t} } \bigg] \nonumber\\ 
P^{\alpha} \left( x , t|* \right) &
= \frac{1}{\sqrt{2 \pi \langle \Delta^{2} x \rangle_{*,t}^{\alpha}} } \exp \bigg[ - \frac{\left( x - \langle x \rangle^{\alpha}_{*,t} \right)^{2}}{2 \langle \Delta^{2} x \rangle_{*,t}^{\alpha} } \bigg]\
\end{align}
it holds that
\begin{equation}
\KL( t |*) = \frac{1}{2} \ln \left[ \frac{\langle \Delta^{2} x \rangle_{*,t}}{\langle \Delta^{2} x \rangle_{*,t}^{\alpha}} \right] + \frac{\langle \Delta^{2} x \rangle_{*,t}^{\alpha} + \left(\langle x \rangle^{\alpha}_{*,t} - \langle x \rangle_{*,t} \right)^{2}}{\langle \Delta^{2} x \rangle_{*,t}} -  \frac{1}{2}
\end{equation}
Again, by choosing the perturbed PDF to describe the system at a time rescaled by a factor $1+ \alpha$,
we find
\begin{equation}
  P^{\alpha}( x,t |*) = P \left( x,\left( 1 + \alpha \right) t |* \right) \implies \langle \O \rangle^{\alpha}_{t} = \langle \O \rangle_{\left( 1 + \alpha \right)t} \approx \langle \O \rangle_{t} + \alpha t \langle \dot{\O} \rangle_{t}
\end{equation}
and
\begin{equation}
  \begin{split}
   \langle x \rangle^{\alpha}_{*,t} 
   & \approx \langle x \rangle_{*,t} + \alpha t \partial_t\langle x \rangle_{*,t} + \frac{\alpha^{2} t^{2}}{2} \frac{\partial^{2}}{\partial t^{2}}\langle x \rangle_{*,t} + \mathcal{O} \left( \alpha^{3}\right)\\
   \langle \Delta^{2} x \rangle^{\alpha}_{*,t} 
   & \approx \langle \Delta^{2} x \rangle_{*,t} + \alpha t \partial_t\langle \Delta^{2} x \rangle_{*,t} + \frac{\alpha^{2} t^{2}}{2} \frac{\partial^{2}}{\partial t^{2}}\langle \Delta^{2} x \rangle_{*,t} + \mathcal{O} \left( \alpha^{3}\right)
  \end{split}
\end{equation}
so that the KL divergence to order $\alpha^{2}$ becomes
\begin{equation}
  \label{eqKLx}
   \KL( t |*) = \frac{\alpha^{2} t^{2}}{2} \left[ \frac{1}{2} \left(\frac{\partial_t\langle \Delta^{2} x \rangle_{*,t}}{\langle \Delta^{2} x \rangle_{*,t}}\right)^2 + \frac{\left( \partial_t\langle x \rangle_{*,t} \right)^{2}}{\langle \Delta^{2} x \rangle_{*,t}} \right] + \mathcal{O} \left( \alpha^{3}\right)
\end{equation}


\section*{References}

\providecommand{\newblock}{}

\end{document}